# A Graph-Partition–Based Scheduling Policy for Heterogeneous Architectures

Hao Wu
FAU Erlangen-Nürnberg
Email: haowu@cs.fau.de

Daniel Lohmann
FAU Erlangen-Nürnberg
Email: lohmann@cs.fau.de

Wolfgang Schröder-Preikschat
FAU Erlangen-Nürnberg
Email: wosch@cs.fau.de

***Abstract*—In order to improve system performance efficiently, a number of systems choose to equip multi-core and many-core processors (such as GPUs). Due to their discrete memory these heterogeneous architectures comprise a distributed system within a computer. A data-flow programming model is attractive in this setting for its ease of expressing concurrency. Programmers only need to define task dependencies without considering how to schedule them on the hardware. However, mapping the resulting task graph onto hardware efficiently remains a challenge. In this paper, we propose a graph-partition scheduling policy for mapping data-flow workloads to heterogeneous hardware. According to our experiments, our graph-partition-based scheduling achieves comparable performance to conventional queue-base approaches.**

## I. Introduction

Since improving system performance by means of promoting a processor's frequency no longer applies today, high-performance systems have adopted multi-core processors with GPGPUs (General-purpose graphics processing units) as accelerators due to their low ratio of power consumption to performance. These heterogeneous systems composed of different processors are used to speed up not only graphics applications, but also general-purpose applications. Although many researchers have reported immense performance boosts using these accelerators, there are still challenges in how to take advantage of these heterogeneous platforms.

In conventional GPGPU programming, programmers consider GPGPUs as acceleration devices and utilize them by uploading input data, launching computation tasks, and moving results back. These usages are not able to provide portability and make GPGPU programming tedious due to manually managing data consistency. In recent years, data-flow programming has been used to address these problems. The data-flow programming model [2], [1] was proposed to ease programming these systems so that programmers use a directed acyclic graph (DAG) to assemble tasks. This programming model is attractive for CPU–GPU and other heterogeneous platforms because a DAG presents the concurrency of a set of tasks without specifying how to schedule this set. In this model, programmers only need to express dependencies between kernels and data, and implement each of these kernels for each type of processor. In this paper, we use the term *kernels* to refer to independent computations and the term *task* for a set of kernels with data dependencies.

Applying this model, the runtime is responsible for scheduling kernels and managing data across processors. Accordingly, a number of scheduling policies are proposed in order to map the data-flow model to heterogeneous hardware. The challenge in this setting is to leverage all types of processors efficiently in order to improve the overall system throughput. In addition, in systems with heterogeneous processors, a system bus connecting both types of processors could be a potential bottleneck [3]. Although unified memory architectures have been proposed by many companies, such as Intel's Haswell, AMD's APU (Accelerated Processing Unit), and NVIDIA's Tegra K1, we focus in this paper on the more general distributed memory architecture with one multi-core CPU and one GPU are connected by a PCI-e bus in the system.

Numerous scheduling policies [1], [2], [7] have been presented to meet these challenges mentioned above. For example, based on performance information including task execution and data transfer, schedulers are able to minimize the makespan of tasks in the ready queue. As data transfer across the system bus could be a nontrivial cost, the data-aware scheduling policy dispatches tasks on the processors where the input data has been placed. Although data-aware scheduling could alleviate the data transfer problem, for complex graph-structured tasks with multiple inputs, this scheduling is not able to dispatch tasks desirably, since input data may have been placed on a different memory node such that costly data transfers across the system bus cannot be avoided.

Another approach to this problem is overlapping task computation and data transfer. However, it is not easy for programmers to choose optimal overlapping policies while requiring extra system resources, such as pinned-memory which might affect other system services. This technique can be used in the graph-partition approach as well.

Today, There is work using graph-partition methods [6], [5] for heterogeneous architectures. However, this policy is only used with one type of processor. For instance, [6] partitions workload on four identical GPUs and employs the work-stealing policy between CPUs and GPUs. We use these methods on processors that have different architectures and performance.

In particular, our contributions are:

- A graph-based scheduler for the data-flow programming model on heterogeneous architectures.
- A comparison between our scheduler and other queue-based schedulers for applications with large graph structures.
- Results showing that this policy is able to reach optimal performance and decrease data transfer frequency in some cases.

## II. Design

By using data-flow programming, users express tasks consisting of kernels for each processor and data dependencies between kernels. In this section, we analyze how to bridge this information to the heterogeneous hardware. In addition, we present requirements for the easy understanding of scheduler behavior.

The goal of our design is to build a system for testing scheduling policies that fulfill the following requirements:

1) The system should provide users with an interface for expressing the graph information of tasks.
2) The scheduler should have the ability to exploit all graph information in order to schedule appropriately.
3) As processors have discrete memory nodes, the system needs to guarantee data consistency.

Regarding the data-flow programming interface, there is a body of previous research that has provided implementations for expressing graph information, either language-based or library-based. In this paper, we use a library-based approach.

In terms of the scheduler policy, it should determine which processor is the most efficient for a kernel and take advantage of computation resources efficiently. First, in order to dispatch kernels to the appropriate processor, it is necessary to acquire characteristics of the tasks' performance on each processor. Second, the scheduler should utilize processors when it is of benefit for the throughput and maintain load-balancing. Third, in order to maintain data consistency, data transfers across PCI-e should be avoided as much as possible when they are particularly costly during the execution time. Based on the analysis above, kernel performance for each processor and data transfers are necessary for desirable scheduling. The methods for acquiring this information includes offline measurements and predictions based on a performance model. In this paper, we use offline measurements.

Finally, in order to understand the scheduler's behavior, the results from the graph-partition should be easily displayed. Also, the configuration for each kernel in one large graph is tedious for programmers. Our system provides methods for configuring kernels in a batch manner to mitigate this work. The implementation will be discussed in the following section.

## III. Implementation

This section describes the implementation of the scheduling framework.

### A. System build

In order to implement a system meeting the requirements discussed above, tools including StarPU, METIS, and DOT are used. StarPU [1] is a unified runtime system for heterogeneous multi-core architectures. StarPU provides a data-flow programming API with data consistency across discrete memory nodes, a number of queue-based schedulers and a plug-in interface that allows users to implement their own scheduling policies. We use StarPU runtime to manage data movement without scheduling. METIS [10] is a graph-partition tool largely used for a distributed system domain. METIS can be applied to partition graphs and finite-element meshes, and to produce fill-reducing orderings for sparse matrices. Its partition functionality for heterogeneous computation nodes is exploited for the CPU–GPU platform. In addition, DOT [12] is a plain text graph description language. It is a simple way of describing graphs that both human and computer programs can use. We use DOT to provide an interface for describing dependencies and to visualize both the original and the partitioned DAGs.

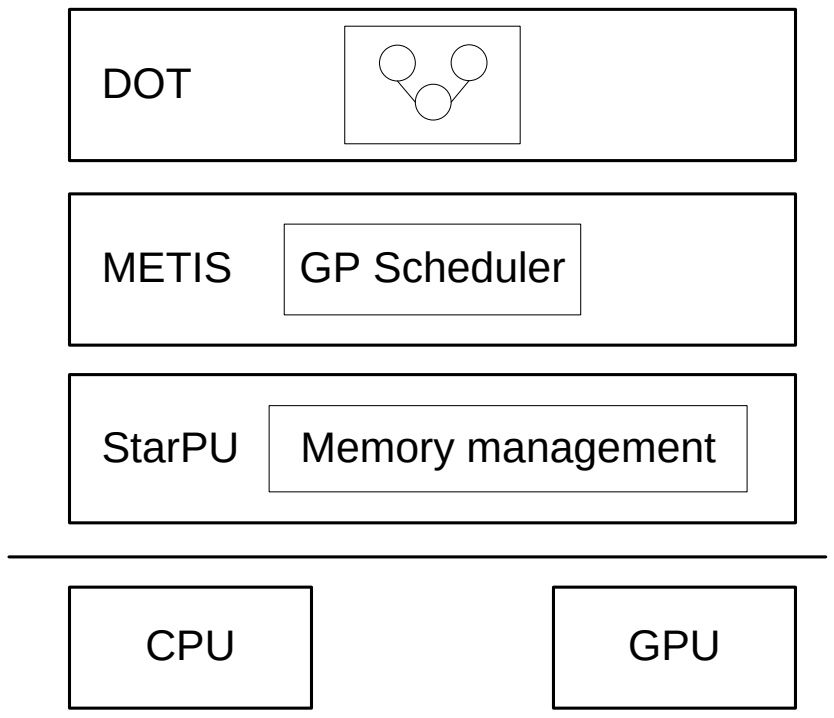

Fig. 1. System Build

As Figure 1 shows, the graph-partition scheduler is implemented on top of the StarPU runtime. The scheduling plug-in interface provided by StarPU does not apply in our case due to its queue-based nature. On top of the scheduler, the DOT is used for describing data dependencies between kernels. Additionally, programmers need to offer kernel implementations for each processor with the StarPU API.

### B. Scheduler framework

Our implementation relies on the StarPU API for expressing kernels and the DOT API for depicting data dependencies. By using a DOT parser and a format translator, DOT, StarPU and METIS can communicate with each other.

When a DAG is provided by programmers, the scheduler first needs to assign a weight value to each edge and each node. These weights for the nodes and edges stand for the performance of each kernel running on the processors and the data transfer overhead, respectively. One approach to acquiring these performance parameters uses a prediction method based on performance models proposed in [13], [14], [15], [16], [17] for multi-core and many-core architectures. Another approach is based on runtime measurements [1], [7]. Due to the limited precision of the performance estimation model, the latter method is applied in this paper to obtain the performance parameters from kernel executions and data transfer costs. The weight values are measured in milliseconds.

Apart from weight information, the scheduler also needs a ratio of workload for each processor. This ratio is calculated based on the performance of kernel execution on each processor and data transfer overhead.

In order to obtain this ratio, we use the formulas

$$R_{CPU} = \frac{Time_{Kernel_GPU}}{(Time_{Kernel_GPU} + Time_{Kernel_CPU})} \quad (1)$$

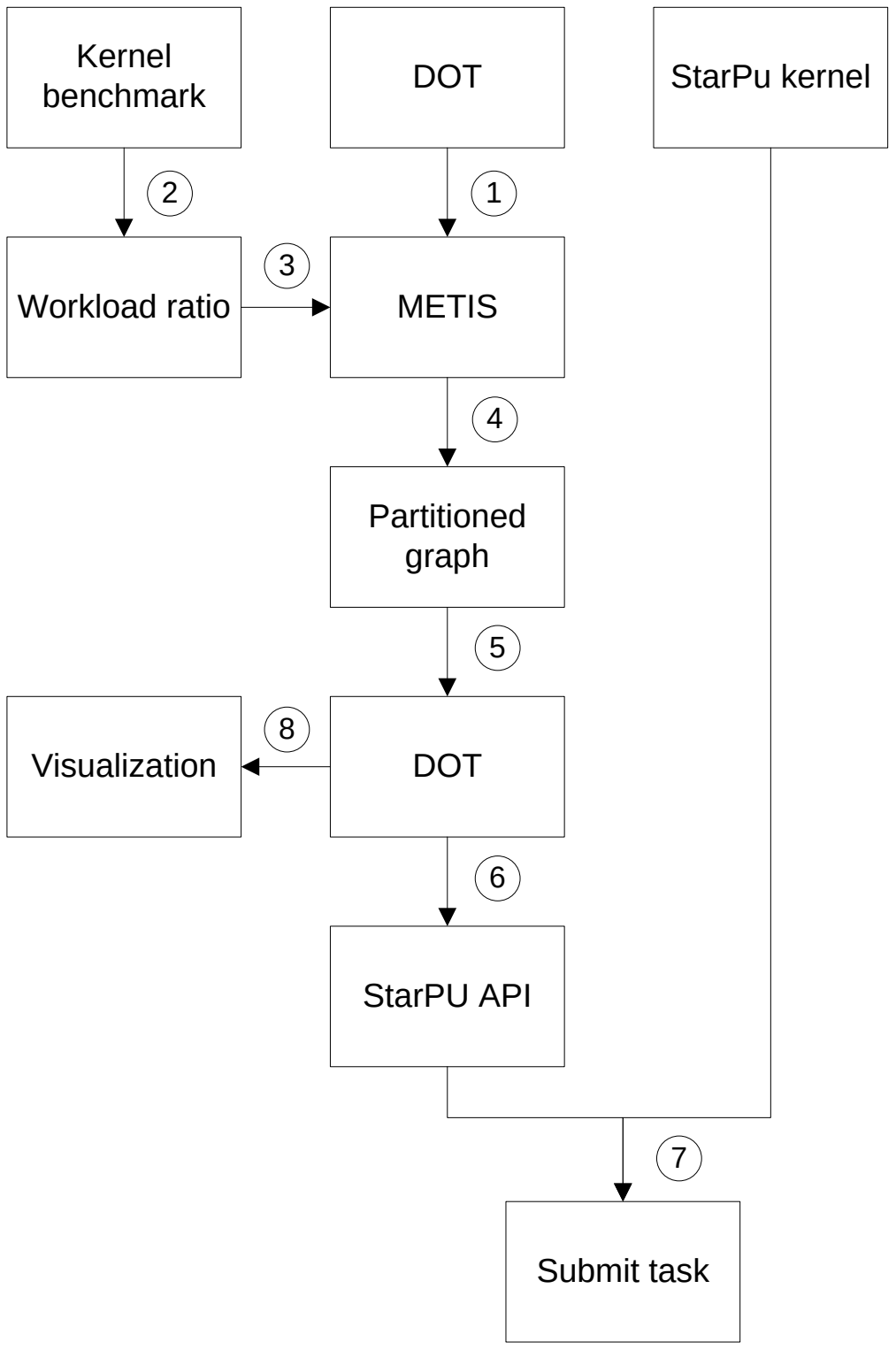

Fig. 2. Processing flow of the scheduler

and

$$R_{GPU} = 1 - R_{CPU}. \quad (2)$$

Another parameter for the partition is the number of partitioned group, which is 2 for the CPU–GPU platform. After partitioning, METIS outputs the partition results. The scheduler translates this result in DOT format and generates the StarPU file. This file contains all kernel and data dependency configurations. As the StarPU API is able to specify on which processor kernels can be scheduled, this feature is used to pin a kernel onto a designated processor. It should be noted that the graph-partition scheduler only pins each kernel onto one processor so that StarPU runtime cannot schedule them again. Finally, with the generated file, including data allocations, kernel configurations and data dependencies, tasks are submitted to the StarPU runtime.

Several issues should be noted when using the graph-partition tool to map the weights of nodes and edges in the graph. First, there are two weight values available for each kernel, i.e., the GPU execution time and the CPU execution time. Choosing either one would not ruin the policy of maintaining load balancing and minimizing data transfer cost. As most applications have a shorter execution time on GPUs than on CPUs, choosing the execution time on GPUs would reduce the node weights. Correspondingly, these small node weights give the edge weights a higher priority during partitioning. On the other hand, choosing the value of CPUs has an opposite effect, which give the edge weights lower priority. How this policy influences the partition results depends on graph partition algorithms. In addition, when mapping the data transfer time for edges, we assume that each transfer with the same size has the same latency no matter whether the direction is from host to device or the reverse. This assumption may not exactly hold on some platform[5]. However, based on measurements on our platform, the error of these two values is within 0.007%, which could be negligible. Finally, compared to GTX GPUs, NVIDIA Tesla GPUs support dual copy engines [8] which allow bi-directional data copy at the same time. This feature can alleviate data transfer overhead. Taking advantaging of this feature will be covered in future work.

TABLE I. EXPERIMENTAL SETUP

| platform | description |
|---|---|
| CPU | 1x Intel Quad-Core i7-4770, 3.40 GHz |
| GPU | 1x GTX TITAN, 2668 cores |
| OS | Linux 3.8.0-29-generic |
| BUS | PCI-e 3.0 16× |

In order to visualize the graph structure, the DOT graph description language is used to express data dependencies between kernels. In the DOT file, an arrow expression refers to a data dependency from one kernel to another. For each kernel, the amount of input equals the number of arrows pointing to this kernel, and the amount of output equals the number of arrows pointing from this kernel. As all initial data is located on the host memory, all initial kernels have data dependencies pointing from an empty kernel whose weight is set to zero.

Figure 2 shows the processing flow of the scheduler. First, programmers provide kernel implementations in the StarPU file and data dependencies in the DOT file. Based on these two files, a complete StarPU file, including data allocation, kernel configurations and kernel dependencies, is generated in order to acquire the performance information of kernels and data transfers. Then, this information is fed to the runtime in order to form a weighted graph and calculate the workload ratio as the input of METIS. After METIS outputs the partitioned result, the final StarPU file, including the scheduling information, is generated. During this process, a format translator is used to bridge METIS and DOT when necessary, since METIS uses a line-based format to express graph rather than a edge-based format used by DOT.

## IV. EVALUATION

To test our approach, this section presents the methodology, the performance characteristics of two selected kernels and comparison of three scheduling policies. Also, the advantages and disadvantages of a graph-partition are discussed.

### A. Methodology

For the experiments, we used a physical machine configured as shown in Table I. In the machine, one quad-core CPU and one GPU are connected via a PCI-e bus. On the CPU side, we used three cores for the workload and one core for the runtime, while on the GPU side we used one worker thread for all GPU workload.

Several possible aspects could affect the performance of an application using the data-flow model. First, the number of kernels and data dependencies determines the structural complexity of this task, which challenges the scheduler with respect to load-balancing and data transfers. We implemented a DAG generator to generate the structure for test tasks.

Specifically, we generated a task with a graph structure including 38 kernels and 75 data dependencies; all kernels are of the same type of matrix computation which has two inputs and one output. In addition, the performance characteristics of kernels could also affect the scheduler behavior. As C.Gregg [3] highlights, the ratio of the execution time on GPUs to the data transfer time plays an important role in the throughput. The higher the ratio is, the more efficient the computation is. Another factor related to performance efficiency is the ratio of the GPU–CPU execution time with the same kind of computation.

We chose the matrices addition (MA) and multiplication (MM) as test kernels due to the differences in their characteristics, which is shown in the following subsection. The set of schedulers we compared consists of: the eager, the data-aware and the graph-partition policies. For each test case, we calculated averages by running 100 iterations in order to minimize the deviation.

### B. The characteristics of kernels

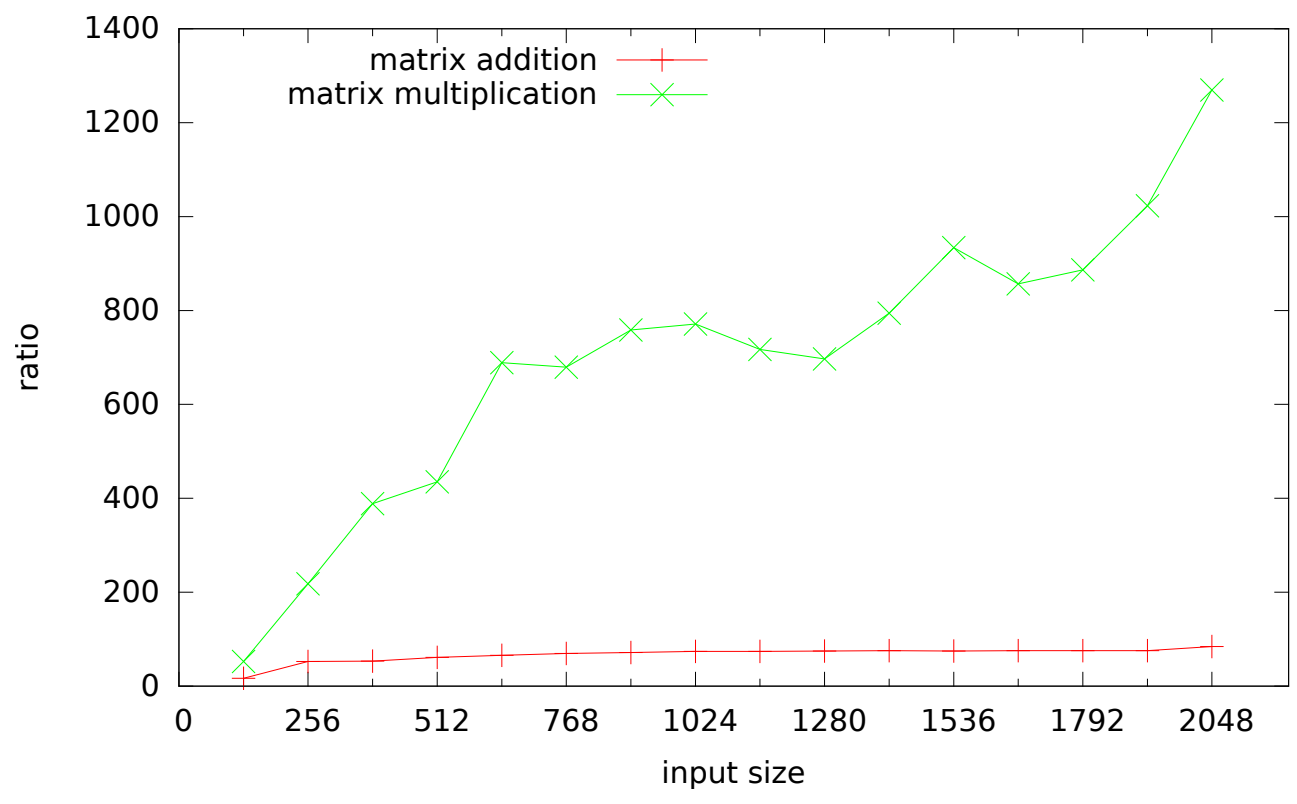

Fig. 3. Ratio of GPU's execution time to CPU

Figure 3 shows the ratio of the execution time on the CPU to the execution time on the GPU. Every matrix is a square-matrix and the x-axis expresses the size of each matrix side. This measurement only considers the computation time on each processor without data transfer. It is shown that the ratio of the MM reflects a steep curve as the input size expands, since computation with exponential complexity can benefit from large-scale parallel cores on GPUs. This indicates that more efficiency can be achieved by scheduling kernels with large input size to GPUs, which is confirmed in the following section. By comparison, the MA kernel maintains a low ratio as the input size increases, which implies that the scheduler may hardly be affected by this kind of computation.

Figure 4 displays the ratio of execution time on the GPU to the time of data transfer across the PCI-e bus. The data transfer includes three matrix transfers with two inputs and one output. The higher the ratio is, the more computation time the kernel needs. The MA with the low curve indicates that it requires the majority of the transferring data, rather than computation. Kernels with this performance characteristic should try to avoid frequent data transfer as much as possible. Meanwhile, the ratio decreases until the size reaches 384 and rises before 1792, then descends again slightly. The reason behind this rise and fall remains unclear. One possible reason could be that the CUBLAS library has optimizations based on the input size.

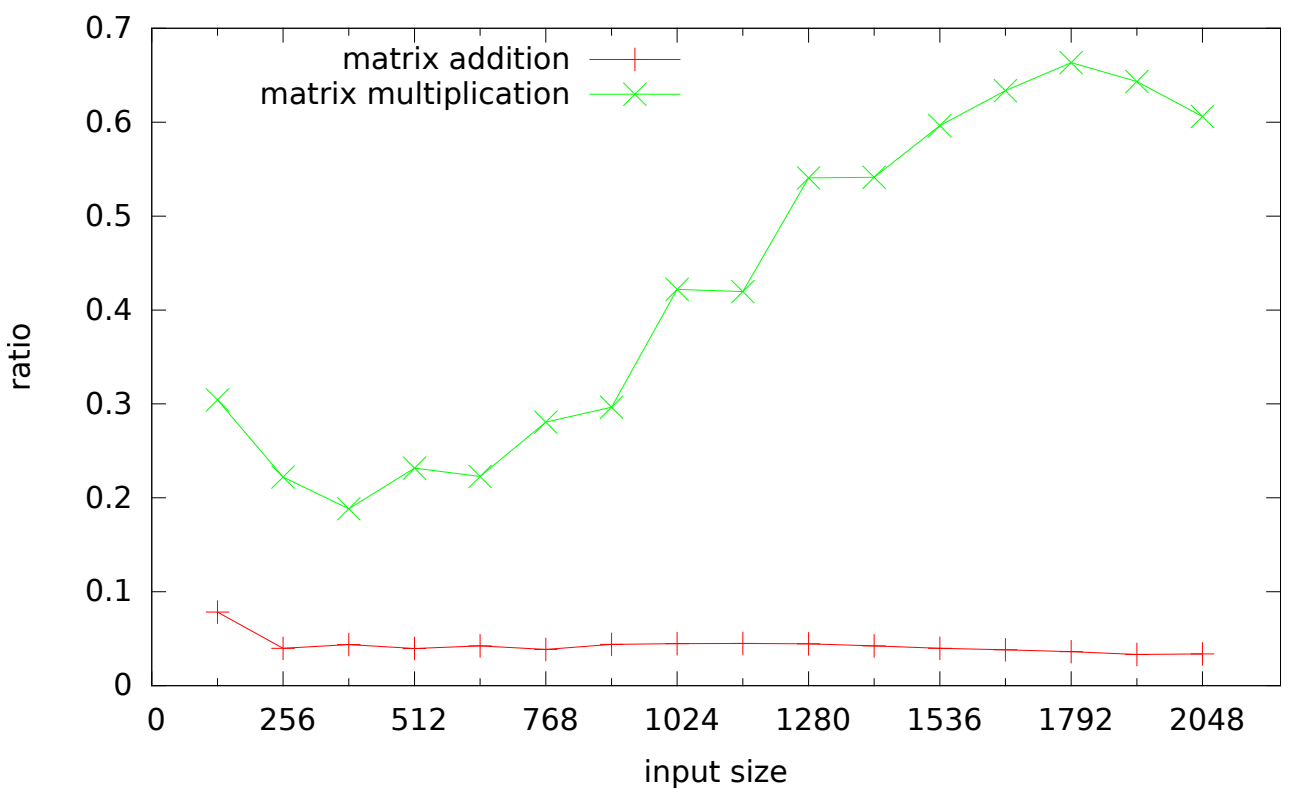

Fig. 4. the ratio of the execution time on the GPU to the data transfer time

### C. Scheduler comparisons

We measured tests tested with two kernels using eager, data-aware (dmda) and graph-partition (gp) scheduling policies. The eager policy tries to exploit both processors when either is idle. The dmda policy tries to schedule kernels on both processors with minimal execution time. The graph-partition policy tries to maintain load-balancing across all processors, and meanwhile, find minimal edge cuts to avoid expensive data transfers.

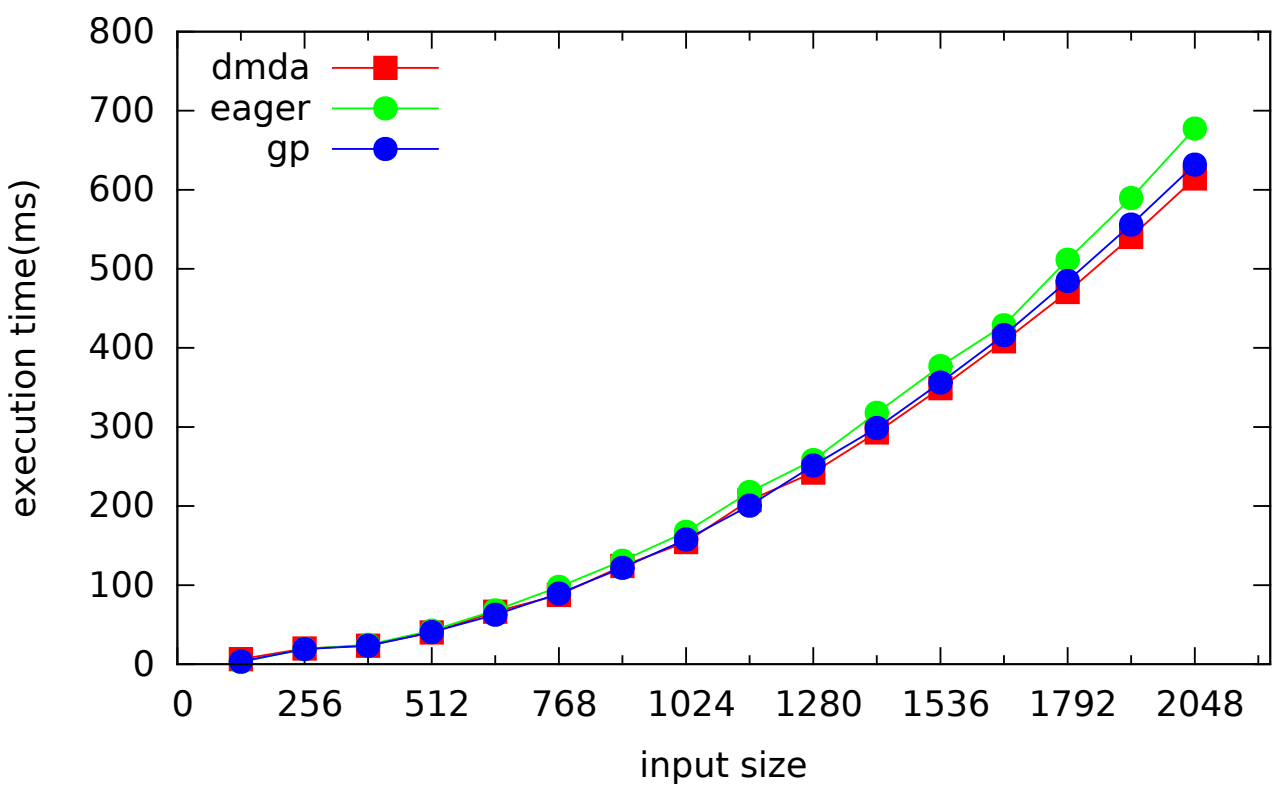

Fig. 5. Execution time for the task with matrix addition kernels

Figure 5 shows the execution time for the task with MA kernels. The performance is close amongst the three scheduling policies. However, the behavior of these policies perform differently. Based on the runtime trace, we found that the eager policy dispatches the most kernels to the GPU and incurs the most data transfer times, as it tries to dispatch kernels on each idle processor and neither considers the total throughput nor the data location. The dmda policy provides less data-transfer times, as the input data location is considered. The graph-partition policy provides the minimal data transfer times. These different scheduling behaviors do not affect the total performance. On the one hand, dispatching this kernel to the GPU cannot attain significant performance, according to the first performance characteristic. On the other hand, exploiting the GPU to share the workload with the CPU incurs large data transfer overhead, according to the second performance characteristic. Therefore, for kernels with these characteristics, GPUs are not efficient for speeding up performance.

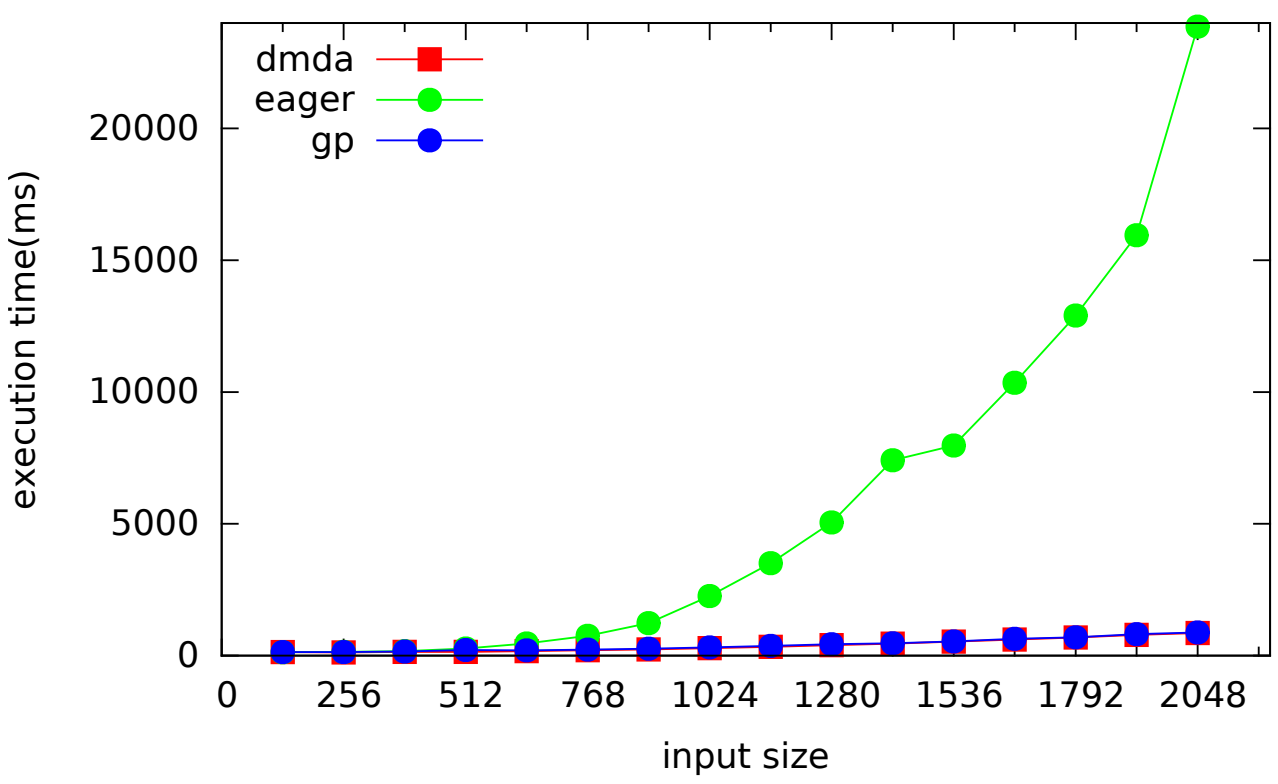

Fig. 6. Execution time for the task with matrix multiplication kernels

Figure 6 displays the performance of the task with the matrix-multiplication kernel. The eager policy shows the highest execution time for all tests and the time increases quickly as the input size expands. As it tries to use each processor without considering the efficiency, some kernels are dispatched to the CPU. For each kernel dispatched to the CPU, its execution time becomes much larger than on the GPU, according to the first performance characteristic. Also, these kernels postpone the execution of the following kernels which depend on their output. By contrast, both the dmda policy and the graph-partition policy achieve a lower execution time, because the dmda policy is aware of the fact that dispatching these kernels on the CPU is not efficient based on the performance history, especially when the input size becomes large. For the graph-partition policy, it makes the same decision to dispatch the entire workload to the GPU for efficiency. More specifically, based on Formula (1) in the previous section, the execution time on the CPU dominates the denominator. Therefore, the workload on the CPU is almost 0, while the workload on the GPU is almost 1. According to this policy, it can be predicted that the CPU could receive a certain amount of workload only when the task largely increases the number of kernels. In this case, as the execution efficiency plays a major role in the throughput, the graph-partition policy does not minimize data transfers. This confirms the idea that if there are large performance gaps between different types of processors, leaving the low-efficiency processor idle can be a better option than using it. Although the graph-partition and the dmda have different scheduling policies, they have the same behavior for the task with this performance characteristic.

### D. *Discussion*

The dmda and the eager policies are online schedulers, while the graph-partition scheduler is currently implemented as an offline scheduler. However, this is an implementation issue and not caused by nature.

The eager and the dmda policies are able to schedule tasks with different performance characteristics. The graph-partition policy assumes that each kernel has the same performance ratio between different type of processors. Hence, we did not test the task consisting of different kernel types on the CPU–GPU platform. This assumption could decrease the generality of this approach. However, this assumption is limited by graph-partition algorithms, not by methods. We believe that applications matching this assumption can benefit from these methods. Graph algorithm researchers may make investigate this assumption in the future. In terms of the scheduling overhead, the dmda policy takes time to make a decision, while the eager does not. The graph-partition scheduler only makes a singular decision and uses the same decision for all following tasks, which averages the scheduling overhead.

## V. Related Work

### A. *Dataflow scheduling*

Scheduling for heterogeneous systems is an active research domain. In the field of scientific computation area, many applications [25], [24] are expressed in a dataflow model. System research such as PTask [2], TimeGraph [20], [23] and others [21] concentrates on eliminating destructive performance interference in the situation of GPU sharing. In addition, Chen et al. [4] proposes a dynamic load-balancing solution for single and multi-GPU systems at a finer granularity. Qilin [22] propose adaptive mapping to map tasks onto the CPU+GPU platform with language support.

Membarth et al. [19] shows an approach to dynamically assigning streams for data-flow programming, which is a CUDA technique, in order to overlap concurrent kernel execution and data transfer.

### B. *Graph partitioning*

Graph partitioning is a common method in distributed system areas. Tanaka et al. [11] presents task assignments for balanced graph-partitioning based on a multi-constraint approach and applies this approach to Montage workflows. The METIS tool also provides a multi-constraint functionality and can be used in our system.

## VI. Conclusions

We proposed a graph-partition scheduling policy for data-flow programming on the CPU–GPU platform. We tested applications with two kernels of different characteristics. Compared with a queue-based scheduler, the graph-partition scheduler was able to achieve a comparable performance and decrease the frequency of data transfer when the performance gap between the processors was not immense. Future work will extend this policy to more heterogeneous systems, such as systems equipped with a CPU, a GPU, and an FPGA and test more applications using the data-flow model.

## Acknowledgment

This work was supported by the Research Training Group 1773 "Heterogeneous Image Systems", funded by the German Research Foundation (DFG).